\documentclass[reprint,showpacs,preprintnumbers,amsmath,amssymb,jcp]{revtex4-1}
\usepackage{graphicx,wasysym}
\usepackage{dcolumn}
\usepackage{bm}
\usepackage{bbm}
\usepackage{natbib}
\usepackage{tabulary}
\usepackage{color}
\usepackage{hhline}
\newcommand{\tobs}{t_N}

\newcommand{\W}{\mathcal{W}}
\newcommand{\Wun}{\mathbb{W}}
\newcommand{\Wt}{\tilde{W}_\lambda}
\newcommand{\app}[1]{\tilde{#1}}
\newcommand{\eigv}{\Xi}

\graphicspath{{./imgs/}}

\usepackage{mathrsfs}

\begin{document}

\title{Constructing Auxiliary Dynamics for Nonequilibrium Stationary States by Variance Minimization}
\author{Ushnish Ray}
\email{uray@caltech.edu}
\affiliation{%
Division of Chemistry and Chemical Engineering, California Institute of Technology, Pasadena, CA 91125, USA
}%
\author{Garnet Kin-Lic Chan}
\email{garnetc@caltech.edu}
\affiliation{%
Division of Chemistry and Chemical Engineering, California Institute of Technology, Pasadena, CA 91125, USA
}%

\date{\today}
\begin{abstract}
  We present a strategy to construct guiding distribution functions (GDFs) based on variance minimization.
  Auxiliary dynamics via GDFs mitigates the exponential growth of variance as a function of bias in Monte Carlo estimators of large deviation functions. 
  The variance minimization technique exploits the exact properties of eigenstates of the tilted operator that defines the biased dynamics in the nonequilibrium system. We demonstrate our techniques in two classes of problems. In the continuum, we show that GDFs can be optimized to study interacting driven diffusive systems where the efficiency is systematically improved by
  incorporating higher correlations into the GDF. On the lattice, we use a correlator product state ansatz to study the 1D WASEP. We show that with modest resources we can capture the features of the susceptibility in large systems that marks the phase transition from uniform transport to a traveling wave state. Our work extends the repertoire of tools available to study nonequilibrium properties in realistic systems. 
\end{abstract}
\maketitle

\section{Introduction}
Large deviation theory (LDT) is a framework to extend the formalism of equilibrium statistical mechanics to nonequilibrium systems~\cite{Touchette2009}.
Much of LDT is concerned with summarizing the dynamics of the system as expressed via the fluctuations of typical trajectories. 
Ensembles of rare trajectories display fascinating behavior reminiscent of phase transitions and criticality. Recent work
illustrates such dynamical behavior in both lattice systems such as simple exclusion processes \cite{Derrida2003, Bodineau2005, Prolhac2009, Jan2011, hurtado2011spontaneous, Gorissen2012, Lazarescu2015, Helms:2019}, constrained kinetic models \cite{nemoto2017finite, Garrahan2019}, models of self-assembly \cite{whitelam2014self, Klymko2018, ray2018a}, dissipative hydrodynamics \cite{Prados:2011, Prados:2012}, and in continuum systems in the form of driven or active Brownian particles \cite{Mehl:2008dw, chetrite2015variational, Nyawo:2016, ray2018a, ray2018b, GrandPre2018}, as well as open quantum systems \cite{Garrahan2018a, Garrahan2018b, Schile2018}. Recently LDT has also been shown to offer a route to calculating nonlinear transport coefficients \cite{Gao2018}.

Accessing  properties of interest, in all but the simplest systems, requires numerical tools. In the context of LDT,
this takes the form of sampling techniques such as the cloning algorithm or diffusion Monte Carlo (DMC) \cite{Grassberger2002, delmoral2005, Giardin2006, giardina2011simulating, Pommier2011, Takahiro2016} and transition path sampling (TPS) or path integral Monte Carlo \cite{bolhuis2002transition};
or representing the non-equilibrium distribution with an explicit ansatz, for example, matrix product states \cite{gorissen2009density, Garrahan2019, Helms:2019}.
The primary challenge in sampling methods is the problem of exponential variance in the estimator
for the large deviation function as a function of the bias.
Several approaches have been suggested to ameliorate this variance problem 
\cite{Klymko2018,Takahiro2016,nemoto2017finite,ray2018b, Whitelam2019}.
These techniques can be interpreted as different forms of importance sampling. 
In previous work, we showed that guiding distribution functions (GDF), as introduced for quantum diffusion Monte Carlo calculations~\cite{Ceperley1980},
define an auxiliary dynamics that  importance samples the dynamics underlying the large deviation function under bias~\cite{ray2018b}, and
highlighted the connection to the generalized Doob's transform~\cite{doobBook, chetrite2015nonequilibrium,sollich2010}. In the current work, we describe a practical numerical technique to generate good guiding distribution functions
in both lattice and continuum simulations of large deviation functions, using the idea of variance minimization. This again draws
from the quantum field, and in particular the methods 
of variational Monte Carlo (VMC)~\cite{Umrigar1988}. There has been other recent work on optimizing auxiliary dynamics, for example in Ref.~\cite{Das2019}; our work provides a different perspective, based on different techniques.



\section{Theory}
\begin{figure*}[t]
\begin{tabular}{cc}
\includegraphics[width=8.6cm]{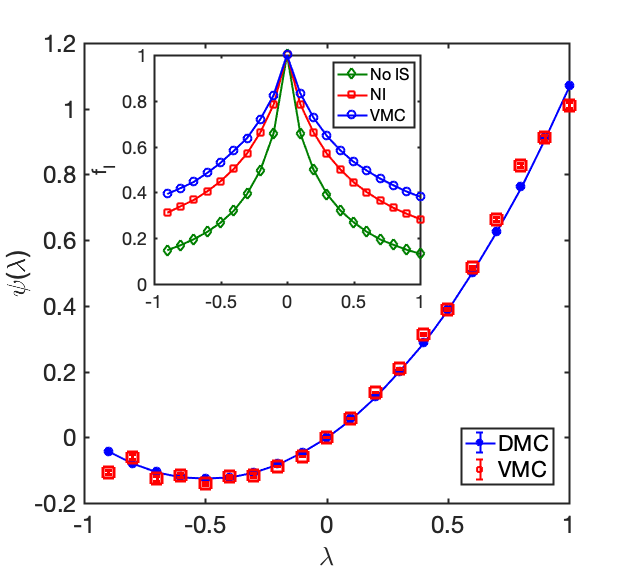} &
\includegraphics[width=8.5cm]{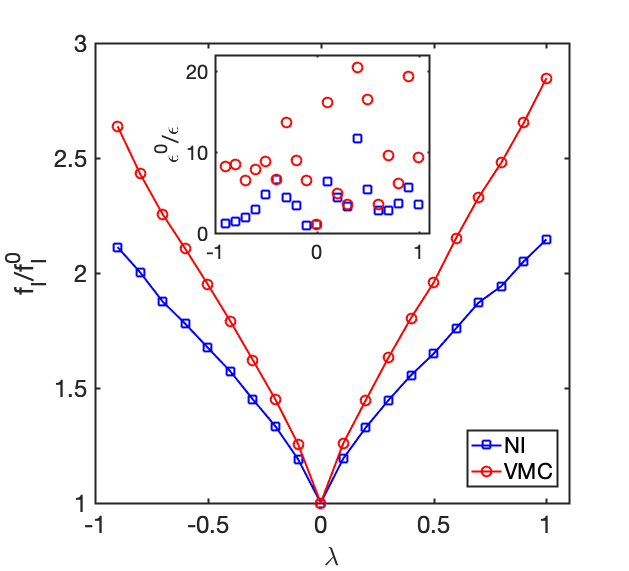} \\
(a) & (b)
\end{tabular}
\vspace{-10pt}
\caption{(a) Large deviation function for the entropy production of $N = 10$ driven Brownian particles in a periodic potential with $v_0=2$, $f=1.0$. The Gaussian interaction force is given by $r_c = 0.10$ and $\alpha = 10.0$. The main figure shows the CGF as a function of $\lambda$ computed using DMC
  using the guiding function, and directly from the GDF via
  the VMC estimator (Eq.~(\ref{eq:vmc_energy})). The DMC calculations are done with different GDF (no GDF, non-interacting (NI) GDF, variational form in main text) but all converge to the same estimate although the efficiencies are different. The inset shows these efficiencies measured via the fraction of independent walkers ($f_I$). 
  (b) Improvement in the fraction of independent walkers ($f_I$) relative to calculations done without any auxiliary dynamics ($f^0_I$) for
  the same 1D Brownian problem. The inset shows the corresponding improvement in the standard deviation of the CGF ($\epsilon(\psi)$). Error bars are smaller than symbol sizes. No error estimate available for the standard deviation (see text). 
}
\label{Fi:1}
\vspace{-10pt} 
\end{figure*}

In the current work, the quantity of interest in large deviation theory is the cumulant generating function (CGF) $\psi(\lambda)$ which is analogous to the free energy of equilibrium statistical mechanics. It is computed via an ensemble average over trajectories given by,
\begin{align}
 \label{Eq:LDF}
\psi(\lambda) &= \lim_{t_N\rightarrow\infty}  \frac{1}{t_N} \ln \left <  e^{\lambda \mathcal{O}[t_N]} \right > \nonumber\\
&=  \lim_{t_N\rightarrow\infty}  \frac{1}{t_N}  \ln \sum_{\mathscr{C}(t_N)} P[\mathscr{C}(t_N)] e^{\lambda \mathcal{O}[t_N]} 
\end{align}
where $\lambda$ is a field conjugate to the observable ${\mathcal{O}} = 
\sum_{t=1}^{t_N} o(\mathcal{C}_{t+},\mathcal{C}_{t-})$, with $o$ an arbitrary function of configurations ($\mathcal{C}_t$) at adjacent times, $t+$ and $t-$, and $t_N$ is the final trajectory time. $P[\mathscr{C}(t_N)]$ is the likelihood of a given trajectory $\mathscr{C}(t_N)=\{\mathcal{C}_0,\mathcal{C}_1,\dots,\mathcal{C}_{t_N}\}$. The trajectories are generated by 
the master equation $\partial_t p_t(\mathcal{C})= \W p_t(\mathcal{C})$, where $p_t(\mathcal{C})$ is the probability of a configuration of the system, $\mathcal{C}$, at time $t$, and $\W$ is a linear operator. It can be shown that $\psi(\lambda)$ is the largest eigenvalue of a ``tilted'' operator $\Wun_\lambda$, i.e., $\langle \eigv| \Wun_\lambda  = \langle\eigv| \psi(\lambda) $ where $\langle\eigv|$ is the corresponding dominant left eigenvector \cite{lebowitz1999gallavotti}. The effect of the tilt is to reweight the transition probabilities of $\W$. 
In the discrete case, $\Wun_\lambda(\mathcal{C},\mathcal{C}') = \W(\mathcal{C},\mathcal{C}')e^{-\lambda o(\mathcal{C},\mathcal{C}')}(1-\delta_{\mathcal{C},\mathcal{C}'}) -R(\mathcal{C})\delta_{\mathcal{C},\mathcal{C}'}$, where $R(\mathcal{C})=\sum_{\mathcal{C}\neq\mathcal{C}'} \W(\mathcal{C},\mathcal{C}')$ is the exit rate.

$\Wun_\lambda$ is not a Markovian operator (i.e. the sum of transition probabilities is not normalized). 
Consequently, when calculating $\psi(\lambda)$ via Monte Carlo techniques it is necessary to track this additional
normalization constant or weight, whose variance grows exponentially with $|\lambda|$. It is desirable thus to
instead consider an auxiliary dynamics generated by a modified operator $\Wt = \hat{\eigv} \Wun_\lambda \hat{\eigv}^{-1}$ (related
to the generalized Doob's transformation~\cite{doobBook, chetrite2015variational}) from which the CGF can be obtained as
\begin{align}
\psi(\lambda) \sim  \frac{1}{t_N} \ln \langle \mathbbm{1}|  \hat{\eigv}^{-1} e^{t_N \Wt}   \hat{\eigv} |p_0\rangle.
\label{eq:DMC2}
\end{align}
where the diagonal matrix $\hat{\eigv} = \sum_{\mathcal{C}} \app{\eigv}(\mathcal{C}) |\mathcal{C}\rangle\langle\mathcal{C}|$ is constructed from the dominant left
eigenvector of $\Wun_\lambda$ (i.e., $\tilde{\Xi}(\mathcal{C}) = \langle\Xi|\mathcal{C}\rangle$),  $\langle \mathbbm{1}|$ is the uniform left vector and $|p_0\rangle$ is the initial distribution of configurations.
The normalization of $\Wt$ is completely independent of configuration and the variance due to the bias is removed, thus
the transformation by $\hat{\eigv}$ carries out a form of importance sampling.
Although propagation with $\Wt$ requires knowledge of the exact eigenvectors, in the GDF approach we simply approximate
these eigenvectors with guiding functions of our own construction and carry out dynamics with $\Wt$ using the diffusion
Monte Carlo algorithm~\cite{ray2018b}.
The quality of the GDF importance sampling
then depends on the degree of overlap between the approximate and the exact left eigenvector of $\Wun_\lambda$.
The problem is therefore reduced to finding a high quality GDF in order to compute the CGF and its associated cumulants with good statistical efficiency.


As mentioned earlier, we can determine GDFs using ideas that originate from quantum diffusion Monte Carlo calculations, where
the analogous problem is to determine a guiding function that best approximates the ground-state of a quantum Hamiltonian~\cite{Umrigar1988}.
This is termed a variational Monte Carlo (VMC) calculation. 
While energy minimization is commonly used for this purpose~\cite{umrigar2007alleviation,sorella2005wave}, $\Wun_\lambda$ is not Hermitian and thus
its spectrum is not necessarily bounded. However, we can use an associated property of eigenstates, \textit{viz.} that the variance of the quantity,
\begin{align}
\Lambda(\mathcal{C}) \equiv \hat{\eigv} \Wun_\lambda |\mathcal{C}\rangle\langle\mathcal{C}|\hat{\eigv}^{-1}
\end{align}
must vanish for eigenstates. This quantity, which we call the \textit{local CGF}, is the analogue of the \textit{local energy} for which
variance minimization has previously been explored in quantum Monte Carlo~\cite{Umrigar1988}. The nonequilibrium variance minimization problem thus
corresponds to minimizing
\begin{align}
\sigma^2(\{p\},\lambda) = \sum_{\mathcal{C}} w(\mathcal{C}) [\Lambda(\{p\},\mathcal{C}) - \langle \Lambda(\{p\})\rangle]^2, 
\label{eq:vmc}
\end{align}
where $ \langle \Lambda(\{p\}\rangle \equiv \sum_{\mathcal{C}} w(\mathcal{C}) \Lambda(\{p\},\mathcal{C})$ is the average of the local CGF, $\{p\}$ are the parameters used to characterize the GDF: $\tilde{\Xi}(\{p\},\mathcal{C})$, and $w(\mathcal{C})$ is a normalized sampling distribution.
The calculation involves minimizing Eq. (\ref{eq:vmc}) with respect to $\{p\}$ over a fixed set of configurations $\{ \mathcal{C}\}$.
Note that (\ref{eq:vmc}) can be sampled from any $w(\mathcal{C})$ but for $w(\mathcal{C}) =  \frac{\tilde{\Xi}(\{p\},\mathcal{C})}{\|\hat{\eigv}(\{p\})\|}$, we obtain
\begin{align}
  \psi(\lambda) \approx \langle \Lambda \rangle = \frac{\langle\tilde{\Xi}(\{p\})|\Wun_\lambda|\mathbbm{1}\rangle}{\langle\tilde{\Xi}(\{p\})|\mathbbm{1}\rangle} \label{eq:vmc_energy}
\end{align}
which is an estimator for $\psi(\lambda)$ in the sense that the approximate sign is replaced by an equality for a GDF that
is the exact eigenstate. Despite the absence of a bounded variational principle, this estimator of $\psi(\lambda)$ 
in practical terms is often useful, and we refer to this as
the VMC estimator for $\psi(\lambda)$, in complete correspondence to its quantum counterpart.

Whereas a strictly zero variance is guaranteed for eigenstates, we emphasize that
a smaller (non-zero) variance of the local CGF does not strictly imply a better GDF.
A more rigorous metric is the actual reduction of the standard deviation in the estimate of the CGF ($\varepsilon(\psi)$) from DMC (or indeed TPS,
which can easily be adapted to use a GDF). Decreasing $\varepsilon(\psi)$ thus defines a better GDF, and this is reflected in statistical independence of samples (the trajectories which are being generated).
For the DMC algorithm the indicator of statistical independence is the fraction of independent walkers $f_I$~\cite{Takahiro2016, ray2018b}.
Empirically $f_I$ is seen to be less susceptible to statistical noise than $\varepsilon(\psi)$ but (for a given $\lambda$) still bears a monotonic relationship with it, and we will therefore primarily use $f_I$ as the measure of quality of a GDF.
In the case of perfect importance sampling, using the exact auxiliary dynamics, $f_I$ is equal to 1 at all times. In the other limit, if all walkers are correlated,  $f_I  = 0$. Because $f_I(t)$ measures the correlation among walkers as a function of time, it must be smallest at $t = 0$ \cite{ray2018a, ray2018b}, which is what we will report. 
It is important to note that although improvements in $f_I$ yield improvements in $\varepsilon(\psi)$ the relationship between the two is not linear. 

\section{Results}

\begin{figure}[t]
\begin{center}
\includegraphics[width=8.5cm]{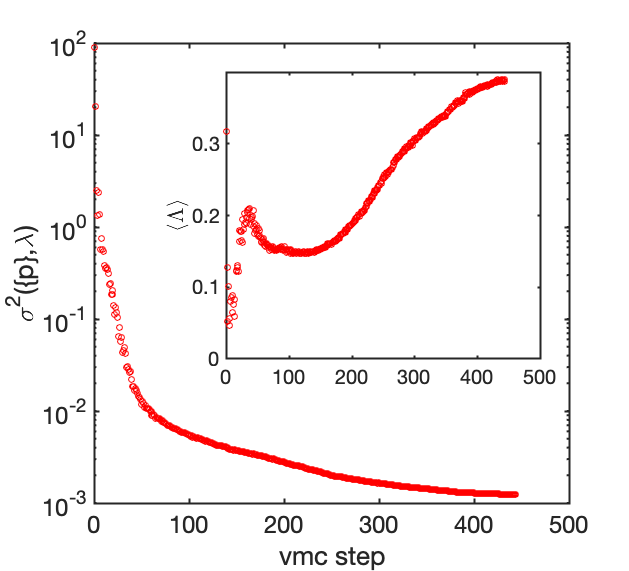}
\caption{VMC minimization trace of the variance for the driven Brownian walker system for $\lambda = 0.5$. Inset shows how the local CGF changes as the
  minimization proceeds (see text).}
\label{Fi:2}
\end{center}
\vspace{-20pt} 
\end{figure}


To demonstrate our procedure, we carry out simulations on a continuum system and a lattice model.
For the continuum system, we consider the prototypical driven Brownian walker, where the observable of interest is the entropy production. This system consists of $N = 10$ particles (at location $\bold{R} = \{r_i\}$) moving in a periodic potential ($V(r) = v_0 \cos(2\pi r)$) on a ring of size $L = 1.0$ under the influence of an external driving force ($f$) and a fluctuating field represented by Gaussian white noise. These particles also interact via a pairwise repulsive force ($\bold{\hat{f}}(r_i,r_j) = - \bold{\hat{f}}(r_j,r_i)$ and $|\bold{\hat{f}}|  = \alpha \exp[-(|r_i-r_j|/r_c)^2]$). The transformed tilted propagator that includes auxiliary dynamics due to the GDF $\app{\eigv}(\bold{R})$ is given by \cite{ray2018b},
\begin{align}
\app{W}_\lambda = &\sum_i \partial_{i} (\partial_{i} - [\bold{F}_i(\bold{R})\cdot\bold{\hat{x}}+2f\lambda + 2\ln \app{\eigv}(\bold{R})]) \nonumber\\
&+ \app{\eigv}^{-1}(\bold{R}) \Wun_\lambda^\dagger \app{\eigv}(\bold{R}),
\end{align}
where $\bold{F}_i(\bold{R}) = f\bold{\hat{x}} - \partial_iV(r_i)\bold{\hat{x}} + \sum_{j\neq i} \bold{\hat{f}}(r_i,r_j)$. The adjoint operator 
${\Wun}_\lambda^\dagger = \sum_i \partial_i^2 + (\bold{F}_i(\bold{R})\cdot\bold{\hat{x}}+2f\lambda)\partial_i + f\lambda(f\lambda+\bold{F}_i(\bold{R})\cdot\bold{\hat{x}})$, represents the norm breaking term that is handled via branching. Trajectories for $\app{W}_\lambda$ are generated from Langevin dynamics $\partial_t{r_i} = \bold{F}_i(\bold{R})\cdot\bold{\hat{x}} + 2f\lambda + 2\partial_i \ln \app{\eigv}(\bold{R})+\eta_i$, where the random force, $\eta_i$, satisfies $\langle \eta_i(t) \rangle=0 \quad$ and $\langle \eta_i(t)\eta_i(t') \rangle = 2 \delta(t-t')$. The entropy production $s(t)$ is
reflected in  the biasing term $\exp [\lambda \tobs s(t)]$ with $s(\tobs) \tobs = \sum_i \int_0^{\tobs} f\dot{\bold{r}_i}(\tau)\mathrm{d}\tau$
which is absorbed into the dynamics of the tilted Fokker-Planck operator~\cite{seifert2012stochastic, ray2018a, ray2018b}.  

\begin{figure*}[t]
\vspace{-18pt}
\begin{tabular}{cc}
\includegraphics[width=8.5cm]{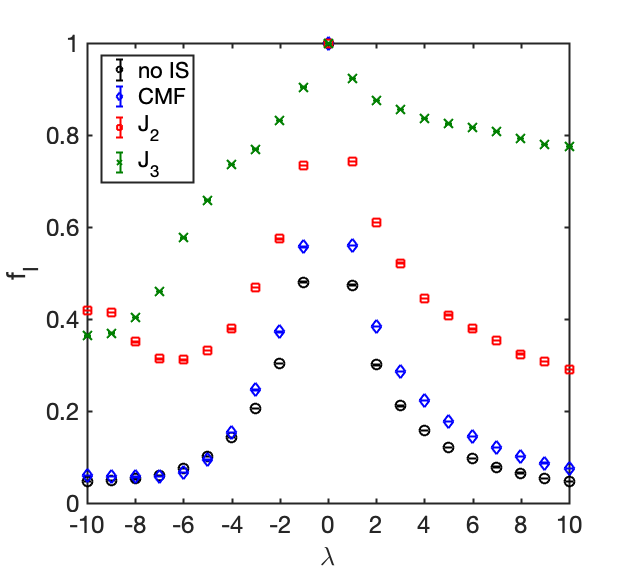} &
\includegraphics[width=8.5cm]{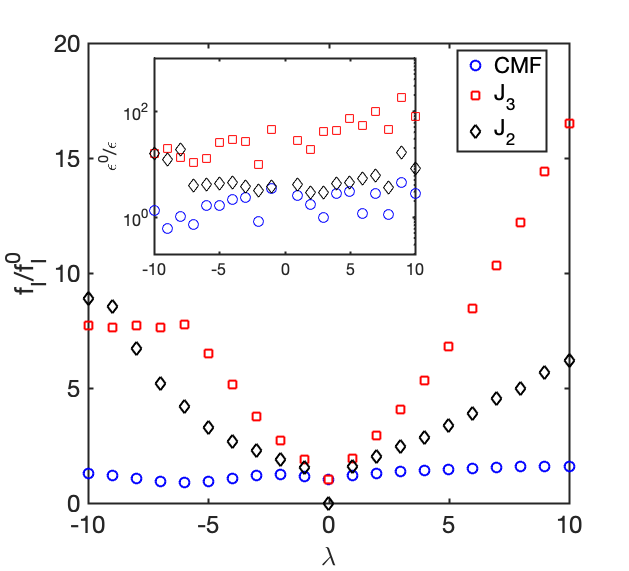}\\
(a) & (b)
\end{tabular} 
\vspace{-10pt}
\caption{(a) Fraction of independent walkers ($f_I$) for different types of GDF (no GDF, non-interacting GDF, optimized variational form (see main text))  for a $L = 16$ WASEP model. The most flexible GDF is provided by the $J_3$ ansatz. We find that the efficiency gains from using the GDF systematically improve from the CMF ansatz to the $J_2$ and $J_3$ ansatz. 
(b) Improvement in the fraction of independent walkers ($f_I$) relative to calculations done without any auxiliary dynamics ($f^0_I$). The inset shows the corresponding improvement in the standard deviation of the CGF ($\epsilon(\psi)$). Error bars are smaller than symbol sizes. No error estimate available for the standard deviation (see text).}
\label{Fi:3}
\vspace{-15pt}
\end{figure*}

For this system we parametrize the GDF as $\tilde{\Xi}(\bold{R}) = \prod_i \phi(r_i) \prod_{j<i} {[J(r_i,r_j)]}$, where the single particle function, $\phi(r_i)$, is obtained from the non-interacting eigenstate of $\app{W}_\lambda$ ($\hat{\mathbf{f}}(r_i,r_j)=0$) generated with $M = 101$ plane wave modes \cite{ray2018b} and $J(r_i,r_j) = \sum_{k_1,k_2} \tilde{J}(k_1,k_2) e^{ik_1r_1 + ik_2r_2}$. The parameters $\tilde{J}(k_1,k_2)$ are estimated by minimizing (\ref{eq:vmc}). 

Shown in Fig.~\ref{Fi:1}a is the large deviation function and $f_I$ computed using no guiding function, the non-interacting guiding
function ($J=1$), and the variational form above (a similar form has been explored independently in \cite{Das2019}).  We parametrized $J(k_1,k_2)$ with $241$ parameters ($21$ plane waves per particle)
and we minimized the variance such that $\sigma^2(\{p\},\lambda) < 4.0\times 10^{-3}$ for all $\lambda$s we considered. The minimization procedure is  started at small $|\lambda|$ using an initial guess for parameters that produces a uniform state since we know that for $\lambda = 0$, the exact $\langle\Xi|$ is uniform. The optimized parameters are used as an initial guess for the next (nearby) $\lambda$.  A trace of the minimization is shown in Fig.~\ref{Fi:2} at $\lambda = -0.5$ obtained using a simple simplex algorithm. For this system the minimization required $10000$ configurations (sampled
from the guess GDF as the distribution $w(\mathcal{C})$) for $-0.3\le \lambda \le 0.3$ ($\lambda \neq 0$) in order to avoid getting stuck in local minima (which produced a poor GDF). For $|\lambda| > 0.3$, $2000$ configurations were sufficient.

From the reduction in $f_I$ (inset of Fig.~\ref{Fi:1}a) it is evident that continuum calculations can be made much more efficient with an appropriate GDFs.
As noted earlier, the improvement in $f_I$ estimated at $t=0$ can imply
greater efficiency gains when the full trajectory space is considered. In Fig.~\ref{Fi:1}b (inset) we show the improvement
in the corresponding standard deviation in the subsequent DMC calculation, which can be reduced by a large factor,
although statistical noise in this measure means that it is difficult to give a precise estimate of the factor.
Additionally, although we used the same observation time ($t_N$) to compute the CGF and cumulants for all types of sampling, we find that the results converge much faster with $t_N$ for simulations done with auxiliary dynamics.\\ 

We now consider an interacting non-equilibrium problem on a lattice, namely the current fluctuations of a periodic weakly asymmetric simple exclusion process (WASEP) \cite{schmittmann1995statistical}. 
The WASEP models transport of $N$ particles on a lattice with $L$ sites. Here $N$ is chosen as $0.3L$. The configuration
of the particles is defined by a set of occupation numbers with hard-core constraints, $n_i=\{0,1\}$, e.g. $\mathcal{C}=\{0,1,..,1,1\}$. 
The tilted propagator for a current bias $\lambda$,  
\begin{align}
\Wun_\lambda = \sum_{i=1}^L &pe^{\lambda/L} \hat{b}^\dagger_{i+1}\hat{b}_i - p \hat{n}_i(\hat{\mathbbm{1}} - \hat{n}_{i+1}) \nonumber\\
+ &qe^{-\lambda/L} \hat{b}^\dagger_{i-1}\hat{b}_i - q \hat{n}_i(\hat{\mathbbm{1}} - \hat{n}_{i-1})
\label{eq:wasep}
\end{align} 
yields particles hopping to the right with rate $p = \frac{1}{2}e^{E/L}$ and to the left with rate $q = \frac{1}{2}e^{-E/L}$, where the $1/L$ factor in hopping gives the weakly asymmetric limit, whose large scale behavior (universality class) is
described by the Edwards-Wilkinson equation~\cite{Prolhac2009,EW1982}.
Here $\hat{b}^\dagger_{i}$ ($\hat{b}_{i}$) creates (destroys) particles on site $i$ and $n_i$ ($\mathbbm{1}-n_i$) counts the number of particles (holes).
For subsequent calculations we set $E = 10$, and use periodic boundary conditions, $i = L+1 \rightarrow 1$ and $i = 0 \rightarrow L$.
\begin{figure*}[t]
\begin{center}
\begin{tabular}{cc}
 \includegraphics[width=8.5cm]{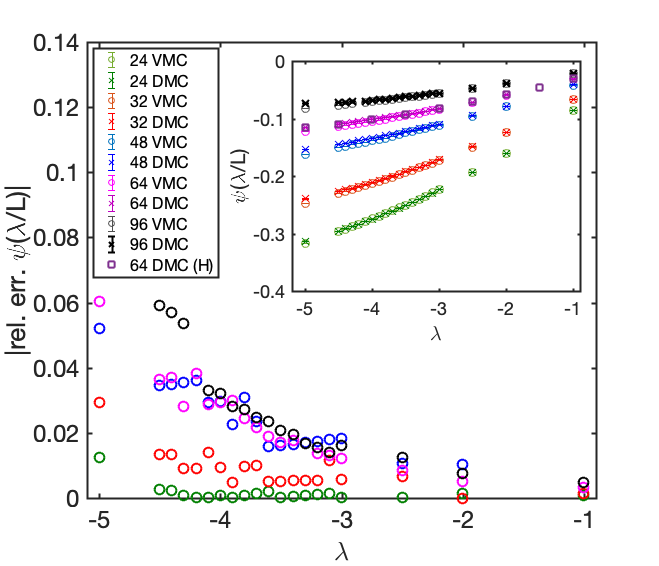} &
\includegraphics[width=8.5cm]{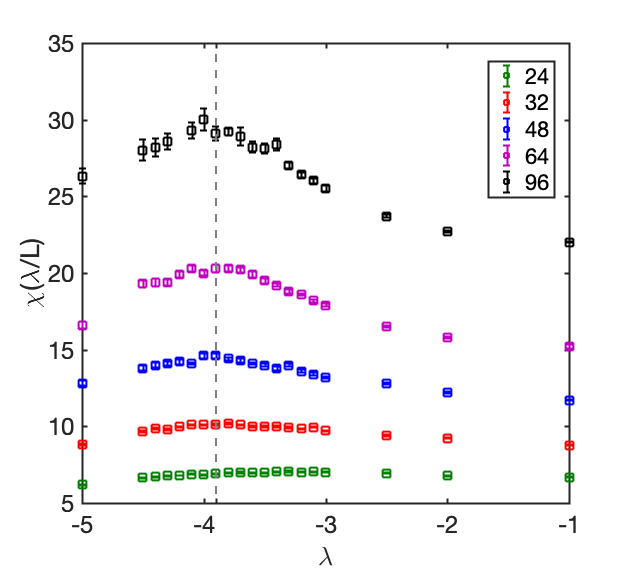}\\
(a) & (b)
\end{tabular}
\vspace{-5pt}
\caption{(a) Relative error of CGF obtained from VMC as compared to DMC. Note that as the system size is increased and more particles are involved the GDF is unable to capture the full extent of the correlations. Nonetheless the DMC calculations still benefit from the GDF in terms of statistical efficiency. Inset shows comparison of DMC and VMC results for the CGF ($\psi(\lambda/L)$). Also included is data from \cite{Hurtado2017} for $L = 64$ indicated as DMC (H).
(b) Susceptibility ($\chi(\lambda/L)$) of the 1D WASEP for different system sizes. The dashed line at $\lambda = 3.9$  indicates the possible location of the continuous phase transition in the limit of $L \rightarrow \infty$.}
\label{Fi:4}
\end{center} 
\vspace{-25pt}
\end{figure*}

Unlike the continuum system, where the soft-core interaction means the non-interacting solution is a sensible starting point to construct the GDF, the hard-core interaction requires a different treatment. Here we consider a GDF that is a product
purely of $n$-particle correlation factors, e.g. for $n=2$, $\tilde{\Xi}(\mathcal{C}) = \prod_{p<q} J_2(r_p,r_q)$ where $r_p, r_q$ denote the
positions of the particles in the configuration $\mathcal{C}$, for $n=3$ we use $J_3(r_p,r_q,r_s)$ etc., and the variational parameters
are the values (i.e. $J_2(r_p,r_q)$). This form is sometimes referred to as a correlator product state in quantum systems~\cite{Nightingale1986, changlani2009approximating}). To enforce PBC we use $J_2(r_{pq}), J_3(r_{pq},r_{qr},r_{qs})$ where the inter-particle
distances are defined with a minimum image convention, i.e. $r_{pq} = \min(|r_p-r_q|, L-|r_p-r_q|)$, and $J_2, J_3$ are symmetric
under cyclic permutations of their arguments. (Cyclic permutation symmetry, rather than full symmetry, was used to reflect the handedness of hopping around in the model). To minimize the number
of variational parameters in the large calculations below, for $r_{pq}>R_\text{cut}$ we set $J_3=1$ where $R_\text{cut}$ is a cutoff distance. For comparison
we have also considered a GDF of the cluster mean-field (CMF) form described in~\cite{ray2018b}.

Fig. \ref{Fi:3}a shows a comparison of $f_I$ as a function of $\lambda$ for a $L = 16$ model with different GDFs. In the WASEP, the short-range CMF is unable to capture the long-range correlations present in the system and therefore we do not get much improvement in efficiency using this GDF. However, with the
correlator product state we can obtain large improvements (e.g., we see from Fig.~\ref{Fi:3}b using $J_3$, $\varepsilon({\psi})$ is improved by an order to two orders of magnitude).

In order to illustrate the flexibility of this ansatz we have further performed calculations for different system sizes $L = 24$-$96$.
Due to the reduction in standard deviation, despite the large system size we needed only a modest number of walkers
in the DMC procedure $N_w = 10000$ to $120000$, which was sufficient to estimate the susceptibility $\chi(\lambda/L) = \frac{d^2\psi(\lambda/L)}{d(\lambda/L)^2}$ (computed as a correlation function).
For these calculations, in order to reduce the number of parameters, we used a cutoff distance of $R_\text{cut}=16$ for $J_3$.
The variance minimization was carried out using between $4000$ to $8000$ fixed configurations sampled from the GDF to
optimize $85$ to $155$ parameters depending on the system size.
The error of the VMC estimator for $\psi(\lambda)$, as a measure of the GDF quality, is shown in Fig. \ref{Fi:4}a where we see that the relative error grows with $|\lambda|$ as more particles become correlated.
Note that for these calculations we did not spend a lot of effort to optimize the more extreme $\lambda$ values as we were interested in the range over which the susceptibility peaks, where the VMC error is $<5\%$.
In this model, the trajectories undergo 
a continuous phase transition from a uniform state to a traveling wave at some $\lambda_c$ provided $E>E_c = \pi/\sqrt{\rho(1-\rho)}$ where $\rho=N/L$  \cite{Bodineau2005, Hurtado2017}. This can be detected from the growing susceptibility of the system shown in Fig.~\ref{Fi:4}b.  Macroscopic Fluctuation Theory (MFT) suggests that $\lambda_c = -E + \sqrt{E^2 - E_c^2}$ $= -2.7198$ (for $\rho = 0.3$)  \cite{Hurtado2017},
although this is not an exact result and
the precise value of $\lambda_c$ for $L\rightarrow\infty$ is not explicitly known \cite{Prolhac2009}. Our DMC calculations suggest a critical point near $\lambda_c \sim 3.9$ for the largest system sizes ($L = 64$, $96$).
We note our results for $\psi(\lambda/L)$ are in excellent agreement with the largest system size (L = $64$) considered in \cite{Hurtado2017}, also plotted in the inset of Fig.~\ref{Fi:4}a.

\section{Conclusions}

In this article we showed how to compute guiding distribution functions
using the technique of variance minimization originating in variational Monte Carlo calculations
of quantum systems. This provides a systematic route to
statistically efficient Monte Carlo computation of large deviation functions using the diffusion Monte Carlo
or cloning algorithm, which we demonstrated in the continuum for the problem of the Brownian walker, as well as on the
lattice for the WASEP model. The very general nature of variance minimization means that the possibilities
for different guiding distribution functions are limited only by one's imagination, and they can be adapted
to very complex systems. Finally, we note that obtaining a good form for the guiding distribution function,
much like obtaining a compact wavefunction in a quantum problem, is likely to provide important analytic insights
into the behaviour of the non-equilibrium system of interest.

\begin{acknowledgements}
  The authors would like to thank Rob Jack, Vivien Lecomte, Juan P. Garrahan and David Limmer for fruitful and engaging discussions. U. R. was supported by the Simons Collaboration on the Many-Electron Problem and the California Institute of Technology. G. K.-L. C. is a Simons Investigator in Theoretical Physics and was supported by the California Institute of Technology and the US Department of Energy, Office of Science via DE-SC0018140.  These calculations were performed with CANSS, available at https://github.com/ushnishray/CANSS.
  \end{acknowledgements}

\appendix
\begin{figure}[t]
\begin{tabular}{c}
\includegraphics[width=8.5cm]{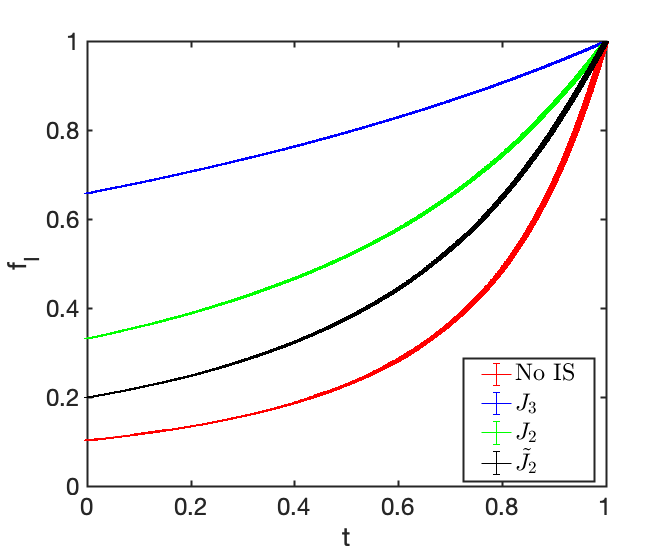}
\end{tabular}
\vspace{-20pt} 
\caption{Fraction of independent walkers ($f_I$) as a function of normalized observation time for different types of auxiliary dynamics for $\lambda = -5.0$. (See text)}
\label{Fi:10}
\end{figure}

\section{Fokker-Planck operator for the continuum}

In the main text we described how the computation of the cumulant generating large deviation function (CGF) 
can be greatly enhanced with the appropriate form of auxiliary dynamics. The GDF structure we have introduced
in the 1D case is of the form $\tilde{\Xi}(\bold{R}) = \prod_i \phi(r_i) \prod_{j<i} {[J(r_i,r_j)]}$.
To carry out simulations with this GDF, we require
the adjoint of the transformed Fokker-Planck operator. For the Brownian walker in the main text, we find
\begin{align}
\frac{\mathbb{W}^\dagger_\lambda \tilde{\Xi}(R)}{\tilde{\Xi}(R)} &= \sum_{i = 1}^N \bigg\{ 
\frac{P_1^\dagger \phi(r_i)}{\phi(r_i)} + \sum_{j\neq i} \frac{\partial^2 J(r_i,r_j)}{\partial r_i^2} + \nonumber\\
&\bigg[\gamma_i(r_i) + 2\lambda + \frac{2}{\phi(r_i)}\frac{d\phi(r_i)}{dr_i} \bigg] \bigg[\sum_{j\neq i} \frac{\partial J(r_i,r_j)}{\partial r_i}\bigg]   + \nonumber\\
&\bigg[ \sum_{j\neq i} \bold{\hat{f}}(r_i,r_j)\cdot \hat{\bold{x}} \bigg]\bigg[ \frac{1}{\phi(r_i)}\frac{d\phi(r_i)}{dr_i} + \sum_{j\neq i} \frac{\partial J(r_i,r_j)}{\partial r_i} \bigg]\nonumber\\
&\bigg\}.
\end{align} 
where the non-interacting single-particle operator is given by,
\begin{align}
\frac{P_1^\dagger \phi(r_i)}{\phi(r_i)} = &\frac{1}{\phi(r_i)}\frac{d\phi(r_i)}{dr_i} + \frac{\gamma_i(r_i)+2\lambda}{\phi(r_i)}\frac{d\phi(r_i)}{dr_i} \nonumber\\
&+ \lambda(\lambda + \gamma_i(r_i))
\end{align}
where $\gamma_i(r_i) = f - \partial_iV(r_i)$ is the effective single-particle force acting on a particle.
This expression provides the norm-breaking term for a subsequent DMC
calculation and  can be generalized straightforwardly to higher dimensions.

\begin{table}[t]
\begin{tabular}{|c|c|c|}
\hline
GDF             & $\psi(\lambda)$ & $\varepsilon(\lambda)$ \\ \hhline{|=|=|=|}
No IS         &  -0.5012     & 1.3e-04 \\ \hline
$\tilde{J}_2$ &  -0.50113   & 9.2e-05 \\ \hline
$J_2$           &  -0.50106   & 2.9e-05 \\ \hline
$J_3$           &  -0.501049 & 4.9e-06 \\ \hline
\end{tabular}
\caption{CGF estimate and standard deviation for different auxiliary dynamics at $\lambda = -0.5$ (see text).}
\label{Tab:1}
\vspace{5pt} 
\end{table}

\section{Statistical Independence and Efficiency Improvement with Auxiliary Dynamics}


In the main text we have mentioned that an improvement of the GDF is indicated by the improvement in the fraction of independent walkers ($f_I$) estimated from DMC (this would be reflected in a corresponding decrease in the autocorrelation time for TPS) and we reported $f_I$ evaluated for $t=0$.
Here we show the full $f_I(t)$ in Fig. \ref{Fi:10} for calculations using different types of dynamics at $\lambda = -5.0$ for the 1D WASEP system.
Note that $f_I(1)=1$ but as one goes back in simulation time, the walkers are descended from a smaller and smaller set of ancestors.
The GDFs $J_2$ and $\tilde{J}_2$ use the same $J_2$ ansatz in the main text but illustrate the effect
of optimizing for more steps in the variance minimization of ${J}_2$, leading to an improved $f_I(t)$.
This is also reflected in Table \ref{Tab:1} where the VMC estimator of the CGF appears to converge with increasingly flexibility
of the GDF, while the standard deviation is systematically improved. 

\bibliography{v1}
\end{document}